\documentclass[aps,prd,preprint,floatfix,nofootinbib,groupedaddress,showpacs]
{revtex4}
\usepackage{graphicx}
\usepackage{amsmath}
\usepackage{latexsym}
\usepackage{here}
\usepackage{array}
\usepackage{dcolumn}
\usepackage{bm}

\def\lsim{\mathrel{\rlap{\raise 2.5pt \hbox{$<$}}\lower 2.5pt
\hbox{$\sim$}}}

\begin{document}
\title{On meson dominance in the `second class'
$\tau\to\eta\pi\nu_\tau$ decay}

\author{N. Paver}
\email[]{nello.paver@ts.infn.it}
\affiliation{Department of Physics, University of Trieste, 34100 Trieste, Italy \\
\&\\
INFN-Sezione di
Trieste, 34100 Trieste, Italy}

\author{Riazuddin}
\email[]{riazuddin@ncp.edu.pk}
\affiliation{Centre for Advanced Mathematics and Physics, National University of
Sciences and Technology, Rawalpindi, Pakistan\\
\&\\
National Centre for Physics, Quaid-i-Azam University,
Islamabad, Pakistan}

\date{\today}

\begin{abstract}
Motivated by recent estimates of the isospin-violating
process $\tau\to\eta\pi\nu_\tau$ in QCD, mostly relying on
the $\rho$ and $a_0$ dominance of the relevant form
factors near threshold, we present an assessment for the
branching ratio that accounts for additional,
potential, effects from the lowest radial excitations
$\rho^\prime\equiv\rho(1450)$ and
$a_0^\prime\equiv a_0(1450)$, respectively, also
lying in the decay phase space.
\end{abstract}


\maketitle
Renewed interest has recently been devoted to the
semileptonic decay $\tau\to\eta\pi\nu_\tau$. This
process belongs to the historical category of
``second class current'' decays \cite{weinberg},
and is only viable in the Standard Model V-A
semileptonic weak interactions through isotopic
spin and G-parity violation, with a probability
strongly suppressed by the smallness of the
$u-d$ current quark mass difference. Therefore,
if measured with a significant branching ratio,
this decay could signal some ``new physics''.
Current theoretical estimates indicate
for this transition a
Standard Model branching ratio of the order of
a unit times $10^{-5}$ \cite{nussinov,neufeld,
pich,truong}, to be compared with the experimental
upper limit $1.4\times 10^{-4}$ \cite{bartelt}. This
should leave some room for revealing ``non-standard physics''
in high statistics studies of this decay, as recently
discussed in Ref.~\cite{nussinov} and, also,
earlier in Ref.~\cite{meurice}.

The theoretical estimates of the
$\tau\to\eta\pi\nu_\tau$ form factors in the
timelike region are based either directly on
{\it ad hoc} $\rho(770)$ and $a_0(980)$ polar
forms or, like in Refs.~\cite{truong,neufeld},
on chiral perturbation theory
\cite{gasser} supplemented, above the $\eta\pi$
threshold, by $\rho$ and $a_0$ poles. In this
note, we present an attempt to
phenomenologically parameterize
the form factors in terms of both the chiral
SU(3)xSU(3) symmetry breaking expansion and
the $\rho,a_0$ resonance poles with, in addition,
account of the lowest $\rho,a_0$ radial excitations
$\rho (1450)$ and $a_0(1450)$. These resonances are
observed experimentally with masses well-within
the $\eta\pi$ invariant mass range allowed to
the decay and, therefore, it should be interesting
to assess the size of their effects relative
to the ground state meson exchanges. Among other
things, this would allow to make tests of
the earlier theoretical predictions.

With $V_\mu={\bar u}\gamma_\mu d$ the weak vector
current, the hadronic matrix element for
$\tau\to\eta\pi\nu_\tau$ can be decomposed into
spin-1 and spin-0 exchange form factors,
respectively, as
\begin{equation}
\label{formfactors}
\langle\pi^+(k)\eta(p)\vert V_\mu \vert 0\rangle =
-\sqrt{2}\left[f_1(t)\left(\left(p-k\right)_\mu -
\frac{M_\eta^2-M_\pi^2}{t}q_\mu\right) +
f_0(t)\frac{M_\eta^2-M_\pi^2}{t}q_\mu\right].
\end{equation}
Here, $q=p+k$ and $t=q^2$ is the invariant mass
squared of the emitted $\eta\pi$ pair. Finiteness of
the matrix element at $t=0$ implies the condition
$f_1(0)=f_0(0)$.

With $t_0=(M_\eta+M_\pi)^2$ and
$\lambda(x,y,z)=x^2+y^2+z^2 -
2(xy+yz+zx)$, the partial width is:
\begin{eqnarray}
\label{decayrate}
\Gamma(\tau^+\to\eta\pi^+\nu_\tau) &=&
\frac{G_F^2\vert V_{ud}\vert^2}{384\pi^3M_\tau^3}
\int_{t_0}^{M_\tau^2}\frac{dt}{t^3}
\lambda^{1/2}(t,M_\eta^2,M_\pi^2)(M_\tau^2-t)^2
\nonumber \\
&\times& \left[\vert f_1(t)\vert^2
\left((2t+M_\tau^2)\lambda(t,M_\eta^2,M_\pi^2)\right) +
\vert f_0(t)\vert^2 3M_\tau^2(M_\eta^2-M_\pi^2)^2\right]
\end{eqnarray}

For the value of form factors at $t=0$,
the prediction from leading and
next-to-leading order in chiral symmetry breaking,
explicitly displaying the isospin breaking character
of the process at the current quark level, is
\cite{truong,neufeld}:
\begin{equation}
\label{chiralimit}
f_1(0)=f_0(0)\equiv\epsilon_{\eta\pi} =
\epsilon_{\eta\pi}^{(2)}+\epsilon_{\eta\pi}^{(4))}; \quad
\epsilon_{\eta\pi}^{(2)}=\frac{\sqrt 3}{4}\frac{m_d-m_u}{m_s-\bar m}.
\end{equation}
In Eq.~(\ref{chiralimit}), $\epsilon_{\eta\pi}$ is
essentially the $\eta-\pi^0$ mixing angle,
and $\bar m$ denotes the average non-strange
current quark mass, ${\bar m}=(m_d+m_u)/2$. The
next-order correction to $\epsilon_{\eta\pi}^{(2)}$ 
is found to be reasonably small, 
$\epsilon^{(4)}_{\eta\pi}/
\epsilon^{(2)}_{\eta\pi}\simeq 0.25$~\cite{neufeld,ecker}.
We include
this correction at $t=0$ in Eq.~(\ref{chiralimit}),
and, consequently, will in the sequel adopt the value
$\epsilon_{\eta\pi}=1.34\times 10^{-2}$, corresponding
to the quark mass ratios $m_u/m_d\simeq 0.55$
and $m_s/m_d\simeq 18.9$ \cite{leutwyler}.\footnote{Also,
isospin breaking of electromagnetic origin will
here be neglected.}

The information on the needed $\rho^\prime$ and
$a_0^\prime$ coupling constants is scarce. For the
radial excitation of the $\rho$, in practice
only $M_{\rho^\prime}\simeq 1.46\ {\rm GeV}$ and
$\Gamma_{\rho^\prime}\simeq 0.4\ {\rm GeV}$ are
found in PDG \cite{pdg}. We model the form factors for
$t\ne 0$ by expressions where
the {\it scale} of the process is set by the
isospin breaking parameter $\epsilon_{\eta\pi}$ as
an overall factor,
times resonant $1^-$ and $0^+$ poles evaluated in the
SU(2) (or the SU(3)) limit. This kind of approximation
will be used for the relevant coupling constants only,
supposedly smooth, but not for the phase space in
Eq.~(\ref{decayrate}).

Accordingly, for the spin-1 form factor we assume,
along the lines of Ref.~\cite{santamaria}, the
unsubtracted polar form with $\rho$ and
$\rho^\prime$
\begin{equation}
\label{spinone-unsubtr}
f_1(t)=\epsilon_{\eta\pi}\times\frac{1}{1+\beta_\rho}\left[
\frac{M_\rho^2}{M_\rho^2-t-iM_\rho\Gamma_\rho(t)} +
\beta_\rho\frac{M_{\rho^\prime}^2}{M_{\rho^\prime}^2-t
-iM_{\rho^\prime}\Gamma_{\rho^\prime}(t)}\right],
\end{equation}
where, denoting by $M$ and $\Gamma$ the resonance
mass and total width, the threshold behaviour of
$\rho,\rho^\prime\to\eta\pi$ is attributed to
$\Gamma(t)$:
\begin{equation}
\label{width}
\Gamma(t)=\theta(t-t_0) \frac{M^2}{t}
\left(\frac{q(t)}{q(M^2)}\right)^{2L+1} \Gamma.
\end{equation}
Here, $L=0,1$ for spin-0 and 1, respectively,
and $q$ is the momentum in the $\eta\pi$ C.M.
frame. A similar parameterization was proposed
in an approach bease on hadron duality and
large number of colors $N_c$, in Ref.~\cite{dominguez}.
Also, in a sense, Eq.~(\ref{spinone-unsubtr}) resembles
the modification of the $\rho$ propagator
introduced in Ref.~\cite{pich1}.

Alternatively, on can consider a once-subtracted
polar form, also satisfying $f_1(0)=\epsilon_{\eta\pi}$:
\begin{equation}
\label{spinone-subtr}
f_1(t)=\epsilon_{\eta\pi}\times\left[1+
t\frac{f_\rho g_{\rho\pi\pi}}{M_\rho^2}\left(
\frac{1}{M_\rho^2-t-iM_\rho\Gamma_\rho(t)} +
\beta_\rho\frac{1}{M_{\rho^\prime}^2-t-
iM_{\rho^\prime}\Gamma_{\rho^\prime}(t)}\right)\right],
\end {equation}

In the above equations, $f$'s and $g$'s are
the vector meson couplings to $e^+e^-$ and to
$\eta\pi$, respectively, for which we have
made the assumption
$g_{\rho(\rho^\prime)\eta\pi}=\epsilon_{\eta\pi}
g_{\rho(\rho^\prime)\pi\pi}$. Numerically,
this determination of the coupling
constant $g_{\rho\eta\pi}$ is qualitatively
consistent with the estimates, obtained by
different approaches, of Refs.~\cite{nussinov} and
\cite{bramon,genz}. The coefficient $\beta_\rho$
in front of the $\rho^\prime$ contribution is
defined as
$\beta_\rho =(M_\rho/M_{\rho^\prime})^2\times
(f_{\rho^\prime}g_{\rho^\prime\pi\pi}/
f_\rho g_{\rho\pi\pi})$.
For a qualitative, theoretical, upper
limit on the $\rho^\prime$ effects, a calculation
within the constituent quark
potential model gives 
$f_{\rho^\prime}/f_\rho\simeq 1.10$
from the ratio of the wave functions at 
the origin, see, e.g., Ref.~\cite{eichten}. One
can assess the value of $g_{\rho^\prime\pi\pi}$
(actually, an upper limit) by identifying
$\Gamma(\rho^\prime\to\pi\pi)$ to
$\Gamma_{\rho^\prime}=0.4\ {\rm GeV}$, so
that the upper limit on $\vert\eta_\rho\vert$ is derived:
$\vert\beta_\rho\vert\leq 0.18$. One can notice that,
for $\beta_\rho\sim -0.17$, the analogues of
Eqs.~(\ref{spinone-unsubtr}) and (\ref{spinone-subtr})
reproduce the branching ratio of about 25\% for
$\tau\to\pi\pi\nu_\tau$ \cite{pdg}, as well as the
fit of the pion for factor from this decay \cite{belle}.
Finally, from experimental data on $\rho$
decays \cite{pdg}, one obtains the ratio
$f_\rho g_{\rho\pi\pi}/M_\rho^2\simeq 1.2$ needed in
Eq.~(\ref{spinone-subtr}), with a small uncertainty.

Apparently, of the two kinds of form factor
parameterisation introduced above,
Eq.~(\ref{spinone-unsubtr}) should be preferable,
as involving a minimal number of input
parameters and being less sensitive to the larger
$t$-behaviour. Nevertheless, also the results from
Eq.~(\ref{spinone-subtr}) will be presented,
as potentially giving some useful information
on the role of the form factor tails and
(upper) limits on the effects due to radial
excitations in the approach followed here.
In Tab.~\ref{tab:spin-one}, we show the predictions for the
spin-1 contributions to the decay branching
fraction, for $\beta_\rho=0$
(no $\rho^\prime$), and for the maximal value
$\vert\beta_\rho\vert=0.18$ obtained above.

\begin{table}[ht]
\begin{center}
\caption{Spin-1 branching ratio vs. $\beta_\rho$.}
\label{tab:spin-one}
\renewcommand{\tabcolsep}{.75em}
\begin{tabular}{|c|c|c|c|}
\hline
BR-spin 1&$\beta_\rho=0$&$\beta_\rho =-0.18$&
$\beta_\rho=0.18$ \\
\hline
Eq.~(\ref{spinone-unsubtr})&$1.69\times 10^{-6}$&
$3.19\times 10^{-6}$&$1.58\times 10^{-6}$ \\
\hline
Eq.~(\ref{spinone-subtr})&$3.97\times 10^{-6}$ &
$3.20\times 10^{-6}$&$5.70\times 10^{-6}$\\
\hline
\end{tabular}
\end{center}
\end{table}

The $\beta=0$ prediction from
Eq.~(\ref{spinone-unsubtr})
is in the ballpark of the earlier estimates, and
the addition of the $\rho^\prime$ could double
the spin-1 branching fraction. Not unexpectedly,
somewhat larger values, still of the same $10^{-6}$
order of magnitude, are obtained from
Eq.~(\ref{spinone-subtr}). One should notice
from Eq.~(\ref{decayrate}) that the
contribution of $f_1(t)$ is suppressed,
relative to $f_0(t)$, by the
behaviour of phase space.

We now turn to the spin-0 form factor $f_0(t)$,
which we assume to be dominated by $a_0$ and
$a_0^\prime$ poles, similar to
Eqs.~(\ref{spinone-unsubtr}) and
(\ref{spinone-subtr}). The nature of
these scalar mesons is not well-established yet,
and we will here assume them to be
$q\bar q$ bound states. From Eq.~(\ref{formfactors}),
$f_0(t)$ is determined by the matrix element
\begin{equation}
\label{divergence}
\langle\pi^+(k)\eta(p)\vert\ i \partial^\mu V_\mu\vert 0\rangle
=(m_d-m_u)\langle\pi^+(k)\eta(p)\vert S^{(1+i2)}\vert 0\rangle
=\sqrt{2}\left(M_\eta^2-M_\pi^2\right)f_0(t),
\end{equation}
with $S^{(1+i2)}={\bar u}d$ the quark bilinear scalar
density. With the definitions
\begin{equation}
\label{fazero}
-\sqrt{2}F_{a_0}M_{a_0}^2=
\langle a_0(q)\vert i\partial^\mu V_\mu\vert 0\rangle,
\qquad
-\sqrt{2}F_{a_0^\prime}M_{a_0^\prime}^2 =
\langle a_0^\prime(q)\vert i\partial^\mu V_\mu\vert 0\rangle,
\end{equation}
we modify the $a_0$ propagator to the unsubtracted polar form
satisfying $f_0(0)=f_1(0)=\epsilon_{\eta\pi}$:
\begin{equation}
\label{spinzero-unsubtr}
f_0(t)=\epsilon_{\eta\pi}\times\frac{1}{1+\beta_a}
\left[\frac{M_{a_0}^2}{M_{a_0}^2-t-iM_{a_0}\Gamma_{a_0}(t)} +
\beta_a \frac{M_{a_0^\prime}^2}
{M_{a_0^\prime}^2-t- iM_{a_0^\prime}\Gamma_{a_0^\prime}(t)} \right],
\end{equation}
and the subtracted version:
\begin{equation}
\label{spinzero-subtr}
f_0(t)=\epsilon_{\eta\pi}\times \left[1+ \frac{t\ f_{a_0}g_{a_0\eta\pi}}{M_\eta^2-M_\pi^2}
\left( \frac{1}{M_{a_0}^2-t-iM_{a_0}\Gamma_{a_0}} + \beta_a
\frac{1}{M_{a_0^\prime}^2-t-iM_{a_0^\prime}\Gamma_{a_0^\prime}(t)}\right)\right].
\end{equation}

Here, the constant $f_{a_0}$ is defined by
$F_{a_0}=\epsilon_{\eta\pi}f_{a_0}$
explicitly showing the isospin violation
of Eq. (\ref{fazero}), and $\Gamma(t)$ is
according to Eq.~(\ref{width}). Moreover,
$g_{a_0\eta\pi}$ and $g_{a_0^\prime\eta\pi}$
are the coupling constants for the decays of
$a_0$ and $a_0^\prime$ into $\eta\pi$,
respectively, and $\beta_a$ is the ratio
$F_{a_0^\prime}g_{a_0^\prime\eta\pi}/F_{a_0}g_{a_0\eta\pi}$.

The coupling for $a_0\to\eta\pi$ has been measured,
with the result $g_{a_0\eta\pi}\simeq 2.80\ {\rm GeV}$
accompanied by a total width
$\Gamma_{a_0}\simeq 100\ {\rm MeV}$
\cite{kloe}, compatibly with~\cite{pdg}. As for the
$g_{a_0^\prime\eta\pi}$ coupling constant, we assume
that the $\eta\pi$, $\eta^\prime\pi$, $K{\bar K}$ and
$\omega\pi\pi$ decay channels of the $a_0^\prime$
saturate the total width
$\Gamma_{a_0^\prime}=265\ {\rm MeV}$.
Using the ratios between partial widths
$\Gamma(\eta^\prime\pi)/\Gamma(\eta\pi)$,
$\Gamma(K\bar K)/\Gamma(\eta\pi)$ and
$\Gamma(\omega\pi\pi)/\Gamma(\eta\pi)$ reported
in \cite{pdg}, one would find
$\Gamma(a_0^\prime\to\eta\pi)\simeq 20\ {\rm MeV}$,
and the consequent estimate
$g_{a_0^\prime\eta\pi}\simeq 1.32\ {\rm GeV}$.
This value is in qualitative agreement with the estimate in
Ref.~\cite{geng} derived from an SU(6) breaking scheme.

To evaluate $\beta_a$, we further need a
theoretical assessment of the ratio
$F_{a_0^\prime}/F_{a_0}$. One possibility
could be to use, along the lines of
Ref.~\cite{nussinov}, current algebra
equal-time commutators in an SU(6) framework
to relate the $0^+$ constant $F_{a_0}$
to the analogous $1^+$ constant $f_{a_1}$,
and for the latter the 2nd Weinberg sum rule
$f_{a_1}=f_\rho$. This would lead to
$-F_{a_0}\simeq (m_d-m_u)f_\rho/M_{a_0}^2$, and
a similar relation for $F_{a_0^\prime}$, so that:
\begin{equation}
\label{su6}
\frac{F_{a_0^\prime}}{F_{a_0}}\simeq
\frac{M_{a_0}^2}{M_{a_0^\prime}^2} \
\frac{f_{\rho^\prime}}{f_\rho}.
\end{equation}
By using the numerical value for the ratio
$f_{\rho^\prime}/f_\rho$ suggested by the previous
arguments made in connection to
Eqs.~(\ref{spinone-unsubtr}) and
(\ref{spinone-subtr}), one would obtain the estimate
$\beta_a\simeq 0.23$.

An alternative is represented by the
``linear dual model'' exploited in the
framework of QCD finite-energy 
sum rules and local hadron
duality in Ref.~\cite{kataev} which, with $S$
denoting $0^+$ `scalar', indicates the rules:
$M_S^2(n)=(n+1)M_S^2$ in perfect agreement
for $n=1$ with the measured $M_{a_0}$, and
$M_{a_0^\prime}$, and $f_S^2(n)=f_S^2/(n+1)$
hence  $F_{a_0^\prime}/F_{a_0}=0.71$. Numerically,
this implies the estimate $\beta_a\simeq 0.33$,
compatible within our approximations with the
previous one.

Finally, for the value of $f_{a_0}$, we
take the QCD sum rule prediction
$F_{a_0}\simeq 1.6\ {\rm MeV}$ \cite{narison}.

These determinations enable us to numerically
exploit Eqs.~(\ref{spinzero-unsubtr}) and
(\ref{spinzero-subtr}) in the assessment of
the spin-0 part of the branching ratio of
interest here. The results, for $\beta_a=0$ and
$\vert\beta_a\vert=0.33$ (the largest value
estimated above), are displayed in
Tab.~\ref{tab:spin-zero}.

\begin{table}[ht]
\begin{center}
\caption{Spin-0 branching ratio vs. $\beta_a$.}
\label{tab:spin-zero}
\renewcommand{\tabcolsep}{.75em}
\begin{tabular}{|c|c|c|c|}
\hline
BR-spin 0&$\beta_a=0$&$\beta_a=-0.33$&
$\beta_a=0.33$ \\
\hline
Eq.~(\ref{spinzero-unsubtr})&$1.07\times 10^{-5}$&
$2.26\times 10^{-5}$&$6.59\times 10^{-6}$ \\
\hline
Eq.~(\ref{spinzero-subtr})&$1.52\times 10^{-5}$ &
$1.51\times 10^{-5}$&$1.56\times 10^{-6}$\\
\hline
\end{tabular}
\end{center}
\end{table}

The size of the spin-0 branching ratio can
be one order of magnitude larger
(${\cal O}(10^{-5})$) than the
spin-1 part. For $\beta_a=0$ (no $a_0^\prime$),
the estimated value is qualitatively consistent
with earlier estimates. Including the $a_0(1450)$
contribution, and depending on the value of
$\beta$, the rate might be increased by a factor of
two.

In conclusion, we confirm the orders of magnitude
of $10^{-6}$ and $10^{-5}$ for the spin-1 and spin-0
branching ratios of previous Standard Model
estimates, these predictions  therefore seem
rather robust. However, the effects from the radial
excitations $\rho(1450)$ and $a_0(1450)$ may be
non-negligible and amount to a factor of two or so,
compared to the simpler $\rho$ and $a_0$ polar forms.
The upper limit from the above tables is
${\rm BR}(\tau\to\eta\pi\nu_\tau)\simeq 2.6\times 10^{-5}$.
On the one side, this value could be in the reach of
future high statistics experiments on $\tau$-decays.
On the other, considering the current
experimental upper limit of order ${\cal O}(10^{-4})$, there
could still remain some room for an unexpected discovery.

\newpage
\leftline{\bf Acknowledgments}
\par\noindent
This research has been partially supported by MiUR
(Italian Ministry of University and Research) and
by funds of the University of Trieste. One of the
authors (R) would like to thank Abdus Salam ICTP
for the hospitality during the summer of 2009,
when this work was started. Our deep thanks to Aqeel
Ahmed and Jamil Aslam for help in the numerical work.

\goodbreak


\begin{thebibliography}{99}

\bibitem{weinberg}
S.~Weinberg, Phys.\ Rev.\ {\bf 112}, 1375 (1958).

\bibitem{nussinov}
S.~Nussinov and A.~Soffer, Phys.\ Rev.\ D {\bf 78},
033006 (2008).

\bibitem{neufeld}
H.~Neufeld and H.~Rupertsberger, Z.\ Phys.\ C
{\bf 68}, 91 (1995).

\bibitem{pich}
A.~Pich, Phys.\ Lett.\ B\ {bf 196}, 561 (1987)

\bibitem{truong}
S.~Tisserant and T.~N.~Truong, Phys.\ Lett.\
{\bf 115B}, 264 (1982).

\bibitem{bartelt}
J.~E.~Bartelt {\it et al.} (CLEO Collaboration),
Phys.\ Rev.\ Lett.\ {bf 76}, 4119 (1996).

\bibitem{meurice}
Y.~Meurice, Phys.\ Rev.\ D {\bf 36}, 2780 (1987);
A.~Bramon, S.~Narison and A. Pich,
Phys.\ Lett.\ {\bf B 196}, 543 (1987).

\bibitem{gasser}
J.~Gasser and H.~Lautwyler, Nucl.\ Phys.\ B\ {\bf 250},
517 (1985).

\bibitem{ecker}
G.~Ecker, G.~M\"uller, H.~Neufeld and A. Pich,
Phys.\ Lett.\ {\bf B 477}, 88 (200).

\bibitem{leutwyler}
H.~Leutwyler, Phys.\ Lett.\ B\ {\bf 378}, 313 (1996).

\bibitem{pdg}
C.~Amsler {\it et al.} (Particle Data Group),
Phys.\ Lett.\ {\bf B 667}, 1 (2008).

\bibitem{santamaria}
J.~H.~K\"uhn and A.~Santamaria, Z.\ Phys.\ C\
{\bf 48}, 445 (1990).

\bibitem{dominguez}
C.~A.~Dominguez, Phys.\ Lett.\ B {\bf 512}, 331
(2001); for a recent application see, e.g.,
C.~Bruch, A.~Khodjamirian and J.~H.~K\"uhn,
Eur.\ Phys.\ J.\ C {\bf 39}, 41 (2005).

\bibitem{pich1}
D.~G\'omez~Dumm, P.~Roig, A.~Pich and J.~Portol\'es,
Phys.\ Lett.\ {\bf B 685}, 158 (2010).

\bibitem{bramon}
Ll.~Ametller and A.~Bramon, Phys.\ Rev.\ D
{\bf 24}, 1325 (1981).

\bibitem{genz} H.~Genz and S.~Tatur,
Phys.\ Rev.\ D {\bf 50}, 3263 (1991).

\bibitem{eichten} E.~Eichten, K.~Gottfried, 
T.~Kinoshita, K.~D.~Lane and T.~M~Yan, 
Phys.\ Rev.\ {\bf 21}, 203 (1980).

\bibitem{belle}
See, for instance, M.~Fujikawa (for the Belle
Collaboration), Nucl.\ Phys.\ B (Proc. Suppl.)
{\bf 169}, 36 (2007); and references there.

\bibitem{kloe}
F.~Ambrosino {\it et al.} (KLOE Collaboration),
Phys.\ Lett. {\bf B 681}, 5 (2009).

\bibitem{geng}
C.~Garcia-Recio, L.~S.~Geng, J.~Nieves and
L.~L.~Salcedo, arXiv:1005.0956 [hep-ph].

\bibitem{kataev}
A.~L.~Kataev,
Phys.\ Atom.\ Nucl.\ {\bf 68}, 567 (2005)
[Yad.\ Fiz.\ {\bf 68}, 597 (2005)].

\bibitem{narison}
S.~Narison, Nuovo Cimento {\bf 10}, 1 (1987);
Nucl.\ Phys.\ Proc.\ Suppl.\ {\bf 186}, 306 (2009).

\end{thebibliography}
\end{document}